\documentclass[preprint]{aastex}

\newcommand{\gtsima}{$\; \buildrel > \over \sim \;$}
\newcommand{\ltsima}{$\; \buildrel < \over \sim \;$}
\newcommand{\simgt}{\lower.5ex\hbox{\gtsima}}
\newcommand{\simlt}{\lower.5ex\hbox{\ltsima}}
\newcommand{\himpc}{{\hbox {$\,h^{-1}$}{\rm Mpc}} }
\newcommand{\himpcz}{{\hbox {$\,h^{-1}{\rm Mpc}_z$}} }
\newcommand{\bfd}{{\mbox{\boldmath $d$}}}
\newcommand{\bfx}{{\mbox{\boldmath $x$}}}

\newcommand{\bfs}{{\mbox{\boldmath $s$}}}

\newcommand{\bfC}{{\mbox{\boldmath $C$}}}
\newcommand{\bfth}{{\mbox{\boldmath $\theta$}}}
\newcommand{\sbfk}{{\mbox{\scriptsize\boldmath $k$}}}
\newcommand{\sbfx}{{\mbox{\scriptsize\boldmath $x$}}}

\newcommand{\omm}{\Omega_{\rm M}}
\newcommand{\oml}{\Omega_{\rm \Lambda}}

\begin{document}



\title{
Constraining the Cosmological Constant from Large-Scale 
Redshift-Space Clustering}

\author{Takahiko Matsubara}
\affil{Department of Physics and Astrophysics, 
	Nagoya University,
	Chikusa, Nagoya 464-8602, Japan}
\and
\author{Alexander S. Szalay}
\affil{Department of Physics and Astronomy, 
        The Johns Hopkins University,
        Baltimore, MD 21218}
\email{taka@a.phys.nagoya-u.ac.jp, szalay@jhu.edu}

\begin{abstract}

We show how the cosmological constant can be estimated from redshift
surveys at different redshifts, using maximum-likelihood techniques.
The apparent redshift-space clustering on large scales (\simgt 20
\himpc) are affected in the radial direction by infall, and curvature
influences the apparent correlations in the transverse direction. The
relative strengths of the two effects will strongly vary with
redshift. Using a simple idealized survey geometry, we compute the
smoothed correlation matrix of the redshift-space correlation
function, and the Fisher matrix for $\omm$ and $\oml$. These represent
the best possible measurement of these parameters given the geometry.
We find that the likelihood contours are turning, according to the
behavior of the angular-diameter distance relation. The clustering
measures from redshift surveys at intermediate-to-high redshifts can
provide a surprisingly tight constraint on $\oml$. We also estimate
confidence contours for real survey geometries, using the SDSS LRG and
QSO surveys as specific examples. We believe that this method will
become a practical tool to constrain the nature of the dark energy.
\end{abstract}



\setcounter{equation}{0}
\section{Introduction}

\label{sec1}

The dark energy, which is the cosmological constant $\Lambda$ in its
simplest form, currently have turned out to possibly be a dominant
component in the universe \citep[e.g.,][]{bah99}. It is one of the
central issues in cosmology to reveal the quantitative nature of this
mysterious form of the energy. One of the mysteries of the
cosmological constant is its smallness. There are no evidence detected
on the Earth for existence of the cosmological constant. The
cosmological constant is so small that it only affects the phenomena
on cosmologically large scales.


There are several traditional tests of the cosmological constant. The
expected frequency of multiple image lensing events for high-redshift
sources is quite sensitive to the cosmological constant
\citep{fuk90,tur90}. Luminosity--volume, and redshift--volume
relations can also be used to measure the geometry of the universe to
constrain the cosmological constant \citep{row68,loh88}. \citet{alc79}
proposed an evolution-free test for the cosmological constant using
statistically spherical objects. The type Ia supernova Hubble diagram
is used to constrain the mass density parameter and the cosmological
constant \citep{sch98,per99} through the luminosity distance $d_L$.
Acoustic peaks of Cosmic Microwave Background (CMB) anisotropies
constrain the curvature of the universe \citep[e.g.,][]{hu97}. Recent
observational developments of the type Ia supernova and the CMB
anisotropies \citep{bal00,deb00} suggest a flat, low-density universe
with positive cosmological constant, $\Omega_{\rm M} \sim 0.3,
\Omega_\Lambda \sim 0.7$.

The effect of a cosmological constant on the clustering properties of
objects in the nearby universe ($z \ll 1$) is so weak, that redshift
surveys have not been used to constrain $\Omega_\Lambda$ so far. As
the depth and the sampling rate of redshift surveys increase,
redshift-space clustering depends on the cosmological constant through
{\em the cosmological redshift distortions}
\citep{bal96,mat96,mat00a}. Several applications of this effect are
proposed \citep{nai99,nak98,pop98,yam99a,yam99b}. To maximally extract
the cosmological information from the survey data, the likelihood
analysis combined with a data reduction technique like the
Karhunen-Lo\`eve transform has been quite successful at low redshifts
\citep{vog96,sza98,mat00b}. We expect it to be just as useful at
intermediate-to-high redshifts here.

In this {\em Letter}, we combine the two methods, i.e., the likelihood
analysis on pixelized data and the cosmological distortions. Specifically, 
we compute the Fisher matrix for simple geometries to illustrate how 
the cosmological distortions constrain the cosmological constant and the 
density parameter in a given redshift survey data.

\section{From Correlations to Fisher Matrix}
\label{sec2}

To generically investigate how a given redshift survey can constrain
the cosmological constant, we construct a rectangular box, in which
objects like galaxies or quasars in redshift space are observed. We
will use smooth pixels rather than hard cells in order to make our
calculations numerically more efficient. We apply a Gaussian smoothing
window to the objects in the survey, to get a smoothed estimate of the
local density. With a sufficiently large smoothing radius, we do not
have to deal with the nonlinearities of the density field. The
Gaussian smoothed cells are placed on lattice sites in the box. In
this way, the smoothed density vector $\rho_i$ on discrete lattice
sites labeled by $i$ is considered as our fundamental data to be
analyzed. In the following we assume the mean value of the density
vector $\langle\rho_i\rangle$ is known so that we can define a
density-fluctuation vector $d_i = \rho_i/\langle\rho_i\rangle - 1$.

In standard theories of structure formation, the linear density field
is a random Gaussian process. In this case, all clustering properties of
the universe are represented by two-point correlations. Thus, a
correlation matrix
\begin{eqnarray}
   C_{ij} = 
   \left\langle
      d_i d_j
   \right\rangle,
\label{eq1}
\end{eqnarray}
theoretically specifies all the statistical information for a given
data set. This matrix is related to a smoothed two-point correlation
function plus a shot noise term:
\begin{eqnarray}
   C_{ij} =
   \int d^3s_1\,d^3s_2\,W(\bfs_i - \bfs_1) W(\bfs_j - \bfs_2)\,
   \xi(\bfs_1,\bfs_2) +
   \int d^3s\,W(\bfs_i - \bfs) W(\bfs_j - \bfs)/\bar{n}(\bfs),
\label{eq2}
\end{eqnarray}
where $W(\bfs) = \exp[-s^2/(2R^2)]/(\sqrt{2\pi} R)$ is a Gaussian
smoothing window, $\xi$ is a two-point correlation function, and
$\bfs_i$ is the position vector of a lattice site $i$, and
$\bar{n}(\bfs)$ is the mean number density field. The position vectors
are all in observable redshift space which consists of the redshift
$z$ and the angular position $(\theta, \phi)$. In the distant-observer
approximation, we can approximately use the Cartesian coordinates in
redshift space and the two-point correlation function is a function of
the relative vector $\bfs_1 - \bfs_2$. The correlation function does
depend on the direction of this relative vector, because of the
redshift distortions. An analytic form of the linear two-point
correlation function in redshift space including high-redshift effects
is given by \citet{mat96}. One of the equivalent forms given in
\citet{mat96} is \citep[see also][]{bal96}
\begin{eqnarray}
   \xi(\bfs_1 - \bfs_2) = 
   b^2(z) D^2(z)
   \int\frac{d^3k}{(2\pi)^3} e^{i\sbfk\cdot(\sbfx_1 - \sbfx_2)}
   \left[1 + \beta(z) {k_3}^2/k^2 \right]^2 P(k),
\label{eq3}
\end{eqnarray}
where $P(k)$ is the linear mass power spectrum at $z=0$, $D(z)$ is the
linear growth rate normalized as $D(0)=1$, $b(z)$ is the bias factor
at redshift $z$, and $\beta(z)$ is the redshift distortion parameter,
which is approximately related to the redshift-dependent mass density
parameter as $\beta(z) \sim {\Omega_{\rm M}}^{0.6}(z)/b(z)$. In the
above equation, the third axis is taken as the direction of the line
of sight. The vectors $\bfx_1$ and $\bfx_2$ are the comoving positions
of the two points which are labeled by $\bfs_1$ and $\bfs_2$ in redshift
space \citep[see][]{mat96}, i.e., they are related by a comoving
distance--redshift relation and the spatial curvature of the universe.
Assuming the distant-observer approximation, a Gaussian window function,
and that the mean number density is effectively constant, $\bar{n}$,
the equation (\ref{eq2}) finally reduces to
\begin{eqnarray}
&&
   C_{ij} =
   \frac12
   b^2(z) D^2(z)
  \int_0^\infty \frac{k^2dk}{2\pi^2} P(k)
   \int_0^\pi \sin\theta d\theta
   \exp\left\{
      - k^2 R^2
      \left[
         \left({c_\Vert}^2 - {c_\bot}^2\right)\cos^2\theta
         + {c_\bot}^2
      \right]
   \right\}
\nonumber\\
&&\qquad\qquad\qquad\qquad\qquad\quad\times\,
   \left(1 + \beta \cos^2\theta\right)^2
   J_0\left(kx\sin\theta_x\sin\theta\right)
   \cos\left(kx\cos\theta_x\cos\theta\right)
\nonumber\\
&& \qquad +\;
   \frac{\exp\left[-x^2/(4R^2)\right]}{\pi^{3/2} (2R)^3 \bar{n}}
\label{eq4}
\end{eqnarray}
In this equation, $c_\Vert(z) = H_0/H(z)$, $c_\bot(z) = H_0 s_K(z)/z$
are the distortion factor parallel and perpendicular to the line of
sight, respectively, where $H(z)$ and $s_K(z)$ are the Hubble
parameter and the comoving angular diameter distance at $z$, and $H_0$
is the Hubble's constant. A line-of-sight component of the
redshift-space distance $s_\Vert$ between the centers of $i$-cell and
$j$-cell is related to that of the comoving distance $x_\Vert$ by
$x_\Vert = c_\Vert(z) s_\Vert$. Similarly, for a component
perpendicular to the line-of-sight we have $x_\bot = c_\bot(z)
s_\bot$. In this notation, the quantities in equation (\ref{eq4}) can
be written as $x \equiv ({c_\Vert}^2 {s_\Vert}^2 + {c_\bot}^2
{s_\bot}^2)^{1/2}$, and $\theta_x = \cos^{-1}(c_\Vert s_\Vert/x)$.
Integration over $\theta$ remains because a spherical Gaussian
smoothing kernel in observable redshift space is no longer spherical
but is ellipsoidal in comoving space. The second term in equation
(\ref{eq4}) is a shot noise term, convolved with the Gaussian kernel.

Once the correlation matrix can be theoretically calculated in any
cosmological model, one can calculate the Cram\'er-Rao bound which
gives an estimate how well the model parameters can be measured.
This is one of the most powerful results in estimation theory
\citep{the92}. The Fisher information matrix is a key quantity in this
theory \citep{ken69}:
\begin{eqnarray}
   F_{\alpha\beta} =
   - \left\langle
      \frac{\partial^2 \ln L}
           {\partial \theta_\alpha \partial \theta_\beta}
   \right\rangle,
\label{eq5}
\end{eqnarray}
where $L(\bfd; \bfth)$ is a probability distribution for the data
vector $\bfd$, which depends on a vector of model parameters $\bfth$.
In our case, the data vector is density fluctuations on lattice sites
and the model parameters are the cosmological parameters. The
Cram\'er-Rao bound states that the maximal likelihood estimate
constrains the model parameters with a minimum variance 
\begin{eqnarray}
   \left\langle
      \theta_\alpha \theta_\beta
   \right\rangle
   \geq
   \left(F^{-1}\right)_{\alpha\beta},
\label{eq6}
\end{eqnarray}
where $F^{-1}$ is the inverse matrix of $F$. When the number of data,
i.e., the dimension of the data vector is very large, the Cram\'er-Rao
bound (\ref{eq6}) becomes equality. A contour $[\theta_\alpha
F_{\alpha\beta} \theta_\beta]^{1/2} = A$ in parameter space gives a
concentration ellipsoid, which indicate regions where the likelihood
density for model parameters are most concentrated. The threshold $A =
1,2,3$ corresponds to maximally attainable confidence levels of
$1\sigma$, $2\sigma$, $3\sigma$, respectively, in a likelihood
analysis, if the likelihood function is Gaussian. The concentration
ellipsoids are useful even when the likelihood function is not
Gaussian to give a rough idea of the spread of the density
\citep{the92}.

As we are interested in a linear density field, the probability
distribution of the density field is considered to be Gaussian so that
the likelihood function has the form,
\begin{eqnarray}
   -2 \ln L = \ln \det \bfC + \bfd^{T} \bfC^{-1} \bfd
   + {\rm const.}
\label{eq7}
\end{eqnarray}
where $\bfC$ is the correlation matrix which depends on model
parameters, and $\bfd$ is the data vector. In this case, the Fisher
information matrix reduces to \citep[see, i.e.,][]{vog96}
\begin{eqnarray}
   F_{\alpha\beta} = \frac12 {\rm Tr}
   \left(
      \bfC^{-1} \bfC_{,\alpha}
      \bfC^{-1} \bfC_{,\beta}
   \right),
\label{eq8}
\end{eqnarray}
where $\bfC_{,\alpha} = \partial\bfC/\partial\theta_\alpha$, etc.
Thus, the Fisher matrix or the concentration ellipsoids for any model
parameters are straightforward to calculate from the correlation
matrix of equation (\ref{eq4}).

\section{Results for a Simple Cubic Box}
\label{sec3}

In this {\em Letter}, the mass density parameter $\omm$ and the
normalized cosmological constant $\oml$, both at the present time, are
the model parameters to be constrained. These parameters has the
primary importance in high-redshift clustering distortions. For
simplicity, the power spectrum $P(k)$ and the bias parameter $b(z)$,
on which the correlation matrix of equation (\ref{eq4}) also depends,
are fixed throughout. We use the cold dark matter-type spectrum with a
fixed shape parameter $\Gamma = 0.2$ and a fixed normalization
$\sigma_8 = 1$. The growth factor $D(z)$ and the distortion parameter
$\beta(z)$ are the functions of $\omm$ and $\oml$. Throughout this
{\em Letter}, we take a fiducial models $(\omm, \oml) = (0.3, 0.7)$ at
which the Fisher matrix is evaluated.

Although the most natural choice of the length unit system would be
comoving coordinate system, we should not use this system, because
they are not actually observable in redshift surveys, and depend on
the cosmological models which we are seeking. Thus, we use the
coordinate system like $(z, \theta, \phi)$ in polar coordinates. The
clustering scale we are interested in is too small in figures when the
distance is represented by $z$ itself, so that we invent a new radial
coordinate $s = cz/H_0 \simeq 2997.9\,z$, i.e., the linear
extrapolation of distance-redshift relation for $z \ll 1$. For
example, a redshift interval $\Delta z = 0.1$ around {\em any}
redshift $z$ corresponds to $\Delta s = 300$. Although the unit of
this coordinate system is still \himpc, we use a new notation \himpcz
to avoid a confusion with the comoving coordinate system.

In this coordinate system, a 200\himpcz cubic box is considered to
obtain generic estimates for Cram\'er-Rao bound, and we compute the
Fisher matrix for this sample, varying the mean redshift $z$ of this
box. The density fluctuations are sampled on regular $10\times
10\times 10$ lattice sites in the box. The Gaussian smoothing radius
is set as $R=10\himpcz$.

\begin{figure}
\epsscale{0.6} \plotone{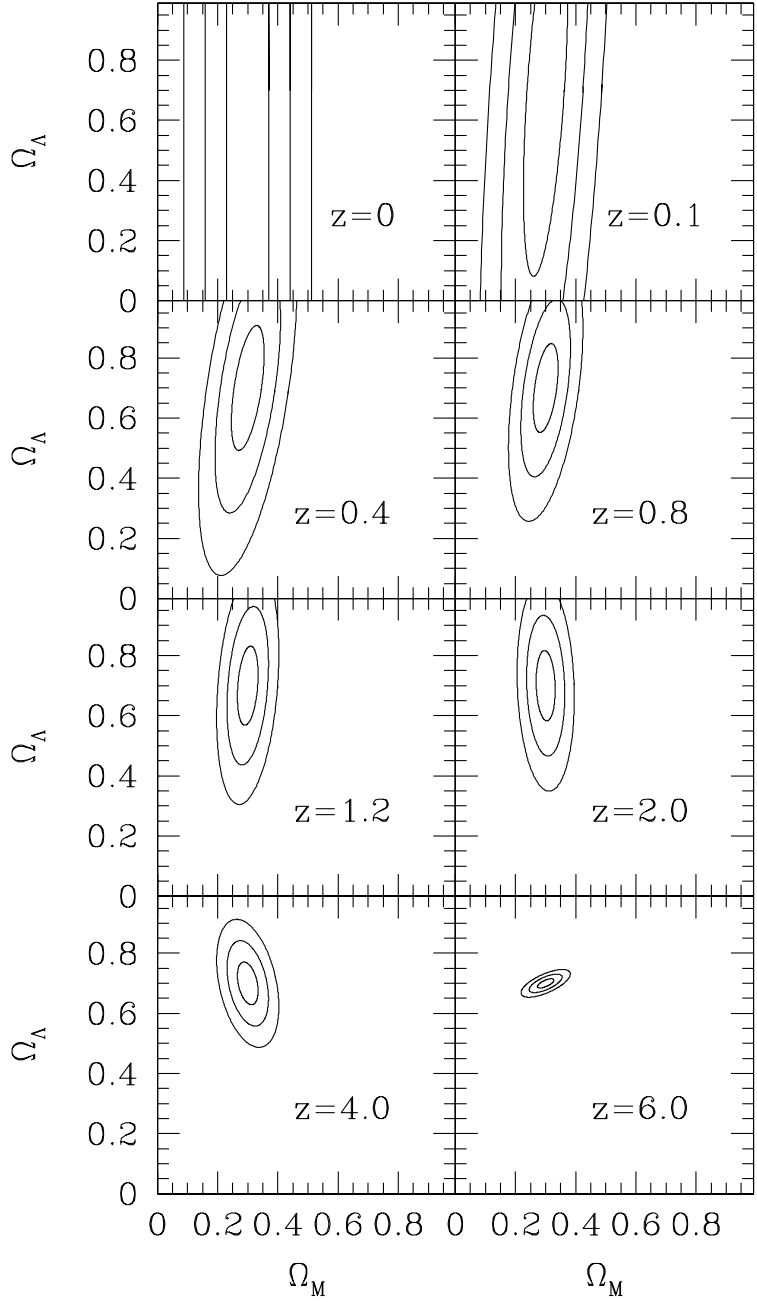} \figcaption[fig1.eps]{Concentration
ellipses from the Fisher matrix for generic boxes. Contour lines
correspond to confidence levels $1\sigma$, $2\sigma$, $3\sigma$
attainable from a single 200\himpcz box placed at redshift $z = 0$ to
$6.0$ as indicated in each panels. This figure contains no shot noise,
and a bias factor of 1.
\label{fig1}}
\end{figure}

Figure~\ref{fig1} shows the concentration ellipses of the 200\himpcz
box placed at redshift $z = 0$ to $6.0$.
The contour
lines correspond to $A=1,2,3$ which indicate the maximally attainable
confidence levels of $1\sigma$, $2\sigma$, $3\sigma$ when one performs
a likelihood analysis for these samples as we described in the
previous section. The shot noise is neglected in this Figure, while it
would be difficult to reduce shot noise for $z\ge 2$ in reality. As is
well known, a low-redshift sample ($z \sim 0$) only constrains the
mass density parameter through its dependence of the redshift
distortion parameter $\beta \sim \omm^{0.6}/b$. Increasing the mean
redshift, the concentration ellipses rotates clockwise and the major
axis becomes shorter, and thus the cosmological constant becomes
constrained. The higher the mean redshift is, the more the
cosmological constant is constrained. Around $z \sim 1.7$, the
concentration ellipses begin to rotate counterclockwise. This is
consistent with the fact that the angular diameter distance--redshift
relation turns over at $z \sim 1.7$ in our fiducial model $(\omm,
\oml) = (0.3, 0.7)$.

\begin{figure}
\epsscale{0.6} \plotone{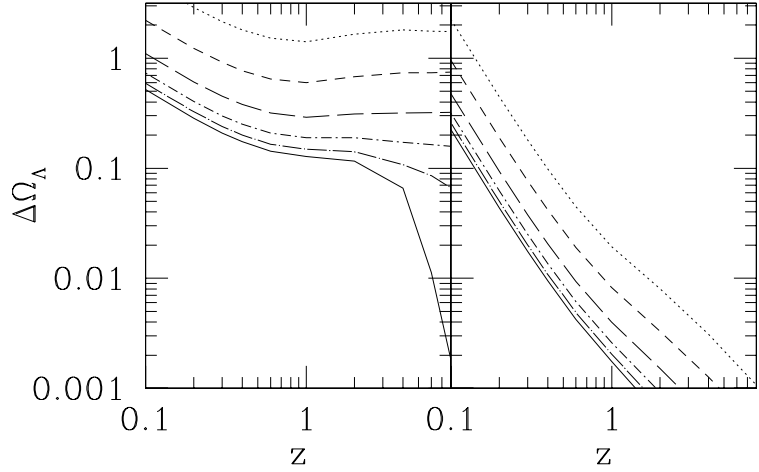} \figcaption[fig2.eps]{Cram\'er-Rao
bound for the normalized cosmological constant $\oml$ as a function of
the mean redshift $z$. Shot noise is varied as $(20\himpcz)^3 \bar{n} =
0.1, 0.3, 1, 3, 10$, and $\infty$, from top to bottom lines. {\em Left
panel}: a single 200\himpcz box. {\em Right panel}: For a survey of
$\pi$ steradians, a redshift interval of $z/2$, centered at $z$. The
number of independent boxes is increased by the volume of the shell,
correspondingly the accuracy improves. A bias factor of 1 has been
used throughout this figure.
\label{fig2}}
\end{figure}

On one hand, the number density one can sample is smaller for
high-redshift objects, which dilute the constraints on cosmological
parameters. On the other hand, there is a larger volume to be sampled
than for low-redshift objects. To obtain the concentration ellipses in
realistic samples, one should take into account both the shot noise
effect and the total volume in a given sample. We again set the
200\himpcz boxes and estimate the Cram\'er-Rao bound with the shot
noise effect included. In Figure~\ref{fig2}, the bound for the
normalized cosmological constant $\oml$ is plotted.
In the left panel, the volume is given by just one 200\himpcz box as
in the case of Figure~\ref{fig1}. The shot noise is varied as
$(20\himpcz)^3 \bar{n} = 0.1, 0.3, 1, 3, 10$, and $\infty$, from top
to bottom lines. In the right panel, the Cram\'er-Rao bound is scaled
by the number of independent 200\himpcz boxes in a $\pi$ steradian
region with a redshift interval $z/2$ around $z$, to obtain a rough
idea of how our error bound is affected by the survey volume. We can
see how densely the objects should be sampled to constrain the
cosmological constant with a certain accuracy both for a sample with a
fixed volume and a sample with a fixed solid angle.

\section{Possibilities for Realistic Survey Volumes}
\label{sec4}

We have considered several different survey layouts for both galaxies
and quasars. The best survey to perform these tests seems to be the
Luminous Red Galaxy (LRG) sample of the Sloan Digital Sky Survey
(SDSS). This sample consists of 100,000 galaxies selected for
spectroscopic observations on the basis of their very red rest frame
colors, using photometric redshifts, down to a limiting magnitude of
$r'=19.5$. They form an approximately volume limited sample, where the
outer edge lies at around $z=0.45$, and the total surface area is
10,000 square degrees.

\begin{figure}
\epsscale{0.4} \plotone{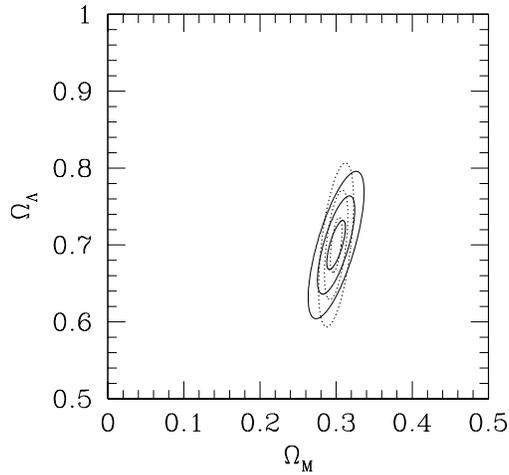} \figcaption[fig3.eps]{Concentration
ellipses corresponding to $1\sigma$, $2\sigma$, $3\sigma$ confidence
levels for approximate geometries of the 100,000 galaxies in the SDSS
LRG (Luminous Red Galaxy) sample. Dotted lines assume a bias factor of
$b=1.5$, solid lines has $b=2$.
\label{fig3}}
\end{figure}

We consider this geometry as a composite of the generic $200\himpcz$
boxes at the mean redshift $z=0.3$. There are about 220 boxes out to
$z=0.45$ in a $\pi$ steradian region, so that the shot noise is
approximately given by $(20\himpcz)^3 \bar{n} = 0.5$. We assume two
bias factors, $b=1.5$ and $b=2$. The resulting concentration ellipses
are shown in Figure~\ref{fig3}.
We can see the result is quite
remarkable. The Cram\'er-Rao bound for the cosmological constant is
only $[(F^{-1})_{\Lambda\Lambda}]^{1/2} = 0.04$ for $b=1.5$, and
$[(F^{-1})_{\Lambda\Lambda}]^{1/2} = 0.03$ for $b=2$. This shows that
the shot noise level and the depth of the survey volume are suitably
balanced to constrain the geometry of the universe in the SDSS LRG
survey.

We have considered various QSO surveys to possibly measure the dark
energy at higher redshifts. Unfortunately, the currently ongoing QSO
redshift surveys, like Sloan Digital Sky Survey \citep[SDSS,
i.e.,][]{yor00} and 2dF QSO redshift survey \citep[2QZ,
i.e.,][]{boy01}, have lower sampling rates for QSOs, $\bar{n} \sim
10^{-3}/(40 \himpcz)^3$. They typically give the Cram\'er-Rao bound of
order $\Delta \Omega_\Lambda \sim 1$, almost regardless of the
smoothing radius. To constrain the cosmological constant with QSO
surveys, one should sample QSOs more densely than these current QSO
surveys. This fact is in agreement with \citet{pop98} who analyzed
nonlinear clustering to constrain the geometry of the universe and
indicated an advantage of a dense sampling.


\section{Discussion}
\label{sec5}

We have shown that large-scale clustering of galaxies at intermediate
redshifts $z \sim 0.5$ is surprisingly suitable for constraining the
cosmological parameters of $\omm$ and $\oml$, and thus the geometry of
the universe. The QSOs in currently ongoing surveys are too sparse to
give comparable constraints.

The apparent redshift-space clustering method used in this {\em
Letter} is a completely self-contained test for $\omm$ and $\oml$. The
results from this method can be further combined with any of the
other independent tests to obtain stricter constraints, or to check a
consistency of our standard picture of the cosmology.

One can use Figure~\ref{fig2} to aid designs of future surveys at
various redshifts. The lines indicate the statistical uncertainty in
the cosmological constant corresponding to different sampling rates.
They should be scaled, noting that the Cram\'er-Rao error bound
roughly scales as the inverse of the square of survey volumes.

In this work, we have only considered two parameters $\omm$ and
$\oml$. We still need to measure the evolution of bias parameter,
which is not obvious. Moreover, there is a possiblity that the dark
energy has a more complex behaviour than the cosmological constant
\citep{wan00}. There are many other cosmological parameters, like
baryonic density $\Omega_{\rm b}$, primordial spectral index $n$, the
neutrino mass density $\Omega_\nu$, etc., which more or less depend on
the apparent redshift-space clustering.

\acknowledgements

AS acknowledges support from grants NSF AST-9802 980 and NASA LTSA
NAG-53503. We would like to acknowledge useful discussions with Dan
VandenBerk, Daniel Eisenstein and Adrian Pope.

\end{document}